\documentclass{emulateapj}
\def\gtsim {\lower .1ex\hbox{\rlap{\raise .6ex\hbox{\hskip .3ex
        {\ifmmode{\scriptscriptstyle >}\else
                {$\scriptscriptstyle >$}\fi}}}
        \kern -.4ex{\ifmmode{\scriptscriptstyle \sim}\else
                {$\scriptscriptstyle\sim$}\fi}}}

\shorttitle{Determining the Nature of Dark Matter} 
\shortauthors{Strigari et~al.}

\begin{document}

\title{Determining the Nature of Dark Matter with Astrometry}  

\author{Louis E. Strigari\altaffilmark{1,2}
 James S. Bullock\altaffilmark{1},
 Manoj Kaplinghat\altaffilmark{1}, 
 \\
}
\altaffiltext{1}{Center for Cosmology,
Dept. of Physics \& Astronomy, University of California, Irvine, CA 92697}
\altaffiltext{2}{McCue Fellow}


\begin{abstract}
We show  that measurements  of  stellar proper motions in dwarf
spheroidal galaxies  provide a  powerful  probe of  the nature of dark
matter.   Allowing  for   general  dark matter density  profiles  and stellar
velocity anisotropy profiles, we  show that the log-slope  of the dark matter
profile at  about twice the stellar core  (King) radius can be
measured  to within $\pm 0.2$ when  the proper motions  of 200 stars are
added  to standard   line-of-sight   velocity dispersion data.    This
measurement  of the log-slope  provides a  test of  Cold and Warm Dark
Matter  theories at  a  sensitivity  not  possible  with line-of-sight
velocity    dispersion   measurements   alone.   The  upcoming  SIM PlanetQuest
 will  have the sensitivity to
obtain the  required   number of  proper motions in   Milky  Way dwarf
spheroidal galaxies.
\end{abstract}




\section{Introduction}  

In a Universe dominated by  collisionless, Cold Dark Matter (CDM), the
central  densities  of galaxy halos are high, and  rise to form cusps with
$\rho \sim r^{-1}$, as $r \rightarrow 0$
\citep{dc91,diemand_etal05}. In contrast, a  
large class of alternatives to CDM, here generically called Warm Dark 
Matter (WDM), are expected to have lower central densities and
constant density cores at small radii
\citep{dalcanton_hogan01,kaplinghat05,cembranos_etal05,Strigari:2006jf}.
In  the  majority of WDM  models, phase  space  arguments predict that
density cores will be more  prominent in lower  mass halos.   
Current observations of the smallest rotationally-supported galaxies
are somewhat ambiguous.  Some galaxies prefer cored halos, some prefer cusps, and others are well fit by either \citep{simon_etal05,kuzio_06}. 
Moreover, the effects of non-circular
motions and other systematic issues are yet to be sorted out for these
systems.

The  dwarf  spheroidal  (dSph)  satellite  galaxies  of the  Milky Way
provide  a potentially superior laboratory  for studying the nature of
dark matter. Observationally,  their proximity ($\sim 100$ kpc) allows
kinematic studies  of  individual stars.  Theoretically,   their small
masses make them ideal candidates  for prominent WDM cores.  Moreover,
the  CDM prediction of cuspy  central  density profiles  is robust for
these   satellite systems because the  cusps  are  stable to any tidal
interactions that may have occurred
\citep{kazantzidis_etal04}.    Unfortunately,  
obtaining dSph density profile slopes has been difficult.
The  best current constraints on
their mass profiles come from analyses of $\sim 200$
 line-of-sight (LOS) velocities \citep[e.g.][]{Lokas2004,Munoz:2005be}.
The stellar velocity anisotropy is a major source of degeneracy in 
LOS velocity dispersion modeling, and, as we show below,
 the log-slope of the
underlying density profile remains practically unconstrained
even if the number of observed stars is increased  by a factor of $\sim 10$.

In  this  {\it Letter}, we  examine  the prospects of constraining the
dark matter   density profiles  of  dSphs  by  combining  stellar  LOS
velocities with  proper motion measurements.   We show that $\sim 200$
proper motions at $\sim 5$ km s$^{-1}$ accuracy can break the relevant
degeneracies   and  determine  the dark    matter distribution in  the
vicinity of the dSph stellar core radius, $r_{\rm king}$. Specifically
both the dark halo  density and the  local log-slope of the dark  halo
density profile  at $\sim 2  \, r_{\rm king}$  may be  determined with
$\gtrsim 5$  times higher precision than  from LOS velocity dispersion
data   alone.  Using two  well-motivated  examples of a  cusp and core
halo,   we show that the local   log-slope measurement can distinguish
between  them  at greater  than the 3-$\sigma$  level. We  discuss our
results  in the context of NASA's  {\em SIM PlanetQuest} (Space Interferometry
Mission) which will have  the  sensitivity to measure proper
motions of stars at this accuracy in multiple dSphs.

Astrometry  as a  means to  constrain  the central density profiles of
dSphs was previously considered  by \citet{Wilkinsonetal02}.  Using a
two-parameter family of models for  the dSphs, these authors show that
proper motions will be enough to completely reconstruct the profile if
the underlying shape follows their  adopted form.  However, CDM halos,
and likely their WDM counterparts, are described by  a minimum of four
unknown parameters:    a  scale density,  a  scale   radius,   an {\em
asymptotic}   inner slope, and an   {\em  asymptotic} outer slope.  By
marginalizing  over  all of these parameters   as well as the velocity
anisotropy of the stars, we demonstrate that the asymptotic slopes are
never well-determined, even with proper motions.  We show that what is
best constrained is the log-slope,  $\gamma(r_*)\equiv - d \ln \rho/ d
\ln r |_{r=r_*}$, at   a radius $r_*$  comparable   to the  King  core
radius, specifically  $r_* \sim  2 \,  r_{\rm  king}$.   We note  that
unlike the asymptotic $r  \rightarrow 0$ slope, the  log-slope at $r_*
\simeq 0.4$ kpc   is known from  CDM  simulations.  This
radius corresponds to  $\sim 1\%$ of the  relevant halo virial radius,
and is well-resolved in current simulations.

\section{Mass Modeling} 

We define the three-dimensional (spatial, not projected) components of
a star's velocity to be $v_r$, $v_\theta$,  and $v_\phi$. The velocity
component along the line of sight is then $v_{los} = v_r \cos
\theta + v_\theta \sin \theta$,  
where $\vec{z} \cdot \vec{r} = \cos \theta$ and $\vec{z}$ is the line-of-sight
direction. 
The components parallel and tangential to the radius vector $\vec{R}$ in
the plane of the sky are $v_R = v_r \sin \theta + v_\theta \cos \theta$ and 
$v_t = v_\phi$, respectively. 
For each component, the velocity dispersion is defined as 
$\sigma_i^2 \equiv \langle v_i^2 \rangle$. 
We will assume $\sigma_\theta^2 = \sigma_\phi^2$. 

The velocity dispersion for each observed component can be constructed
by solving the  Jeans equation for the 
three-dimensional stellar radial velocity dispersion
profile $\sigma_r(r)$ 
and integrating  along the line of sight.
We note that even in the case of tidally disturbed
dwarfs, \citet{k_etal06} have shown that 
 dSph velocity dispersions are well modeled  by  the Jeans
equation,  as long as unbound, interloper stars are removed with standard
procedures. We derive the three resulting observable velocity
dispersions:
\begin{eqnarray}
\sigma_{los}^2(R) & = &  \frac{2}{I_\star(R)} \int_{R}^{\infty} \left ( 1 - \beta \frac{R^{2}}{r^2} \right )
\frac{\nu_{\star} \sigma_{r}^{2} r dr}{\sqrt{r^2-R^2}} \,, \label{eq:LOSdispersion} \\
\sigma_R^2(R) & =  & \frac{2}{I_\star(R)} \int_{R}^{\infty} \left ( 1
- \beta+ \beta \frac{R^{2}}{r^2} \right )
\frac{\nu_{\star} \sigma_{r}^{2} r dr}{\sqrt{r^2-R^2}} \,, \label{eq:Rdispersion}\\
\sigma_t^2(R) &  = & \frac{2}{I_\star(R)} \int_{R}^{\infty} \left ( 1 - \beta \right )
\frac{\nu_{\star} \sigma_{r}^{2} r dr}{\sqrt{r^2-R^2}} \,. \label{eq:phidispersion} \\
\nonumber
\end{eqnarray}
Here $\beta(r) = 1 - \sigma_\theta^2/\sigma_r^2$ is the stellar velocity
anisotropy,  $I_\star(R)$ is the surface density of stars, and $\nu_*(r)$ is 
the three-dimensional  density of stars. 
It is clear from inspection that each component depends on $\beta$ in a different
fashion, and therefore can be used together to constrain its value.
For $I_\star(R)$ and
$\nu_\star(r)$ we use a King profile  \citep{King62}, which is
characterized by a core radius, $r_{{\rm king}}$, and tidal 
radius,  $r_t$. We  adopt values that describe the surface
density of Draco: $r_t = 0.93$ kpc and $r_{\rm king} =0.18$ kpc.  Note that the remaining dSphs have similar
King concentrations, $r_t/r_{\rm king} \sim 5$, with Sextans having  the largest ratio
$r_t/r_{\rm king} \sim 10$. 
Our results do not change significantly as we vary
$r_t/r_{\rm king}$ (equivalent to looking at different dSphs).

In Eqs.~(\ref{eq:LOSdispersion}), (\ref{eq:Rdispersion}) and
(\ref{eq:phidispersion}),  
the radial stellar velocity dispersion, $\sigma_r$, 
depends on the total mass distribution, and thus the parameters
describing the dark matter density profile. We will consider the following
general parameterization of the dark matter density profile, 
\begin{equation}
\rho (r) = \frac{\rho_0}{(r/r_0)^a [1+(r/r_0)^b]^{(c-a)/b}}. 
\label{eq:densityprofile}
\end{equation}
Here, the value of $a$ sets the {\em asymptotic}
inner slope, and different combinations
of $b$ and $c$ set the transition to the asymptotic outer slope. 
For the specific choice $(a,b,c) = (1,1,3)$, we have an NFW profile
\citep{nfw96}.  We denote this as our {\em cusp} case below.
For our {\em core} case we use 
$(a,b,c)=(0,1.5,3)$, corresponding to a  Burkert profile
\citep{burkert95}. We take these two models to be
representative of the predictions of CDM and WDM models.
The Burkert profile for the core case is motivated by the
expectation that  WDM halos will mimic CDM halos at large radius.
This was seen in the WDM simulations of
 \citet{colin_00}.
The Burkert choice is also conservative compared to the often-used
isothermal core with $(a,b,c) = (0,2,2)$, which is more divergent in
shape from an NFW and would be
easier to distinguish observationally.  Regardless, 
our methods are robust to changes in the underlying form of the
density profile.
 
Eq.~(\ref{eq:densityprofile}) allows considerable flexibility
in overall form, and the five {\em shape} parameters ($a,b,c,r_0,\rho_0$)
are in many cases degenerate.  However, there are a number of
physically relevant quantities that may be 
 derived for any set of the five shape parameters.
The first is the log-slope of the dark matter density profile,  defined as 
 $\gamma(r) = - d \ln \rho(r) /d \ln r$. For the density
 profile in Eq.~(\ref{eq:densityprofile}) this is given by
$\gamma (r) = a-(a-c)(r/r_0)^b/[1+(r/r_0)^b]$.
Other quantities of physical interest are the integrated mass within a given radius, $M(r)$, and the physical density at a given radius, $\rho(r)$, which
are clearly obtained for a degenerate set of shape parameters.
Below, we show that while the {\em shape} parameters are not well constrained by dSph velocity data, the {\em physical} quantities of interest
at the scale of the stellar core radius, $r_\star \simeq 2 \, r_{\rm king}$, 
 may be constrained to high precision.

\section{Forecasting Errors on Parameters}
Our goal is to estimate the accuracy with which 
the velocity components of stars in dSphs can be used to 
probe the underlying dark matter distribution. We will
consider a model with six independent parameters: $a$, $b$, $c$, $\rho_0$, 
$r_s$, and $\beta= $ constant.  
We will   
consider generalized $\beta(r)$ forms below.
In order to keep the profile shape relatively smooth (as is expected
for dark matter halo profiles) we restrict the range of $b$ and $c$ by
adding Gaussian priors of $\pm 2$.
  
The errors
attainable on these parameters will depend  on the covariance matrix,
which we will approximate by the $6 \times 6$  Fisher information
matrix  
$F_{\imath \jmath} = \langle \partial^2 \ln {\cal L} /
\partial p_\imath \partial p_\jmath  \rangle $ \citep{kendallstuart69}. 
The inverse of the Fisher matrix, ${\bf F}^{-1}$, 
provides an estimate of the covariance between the parameters, 
and $\sqrt{F_{\imath\imath}^{-1}}$ approximates the error in the
estimate on the parameter $p_\imath$. The Cramer-Rao inequality
guarantees  that $\sqrt{F_{\imath\imath}^{-1}}$ is the minimum
possible variance on the $\imath$th  parameter for an unbiased
estimator. 
Using ${\bf F}^{-1}$ in place of the true covariance matrix involves
approximating the likelihood function of the parameters as Gaussian
near its peak, so  ${\bf F}^{-1}$ will be a good approximation
to the errors on parameters that are well-constrained. The Fisher
matrix also provides information about degeneracies between parameters
but obviously should not be trusted for estimates of the error along
these degeneracy directions.

\begin{deluxetable*}{lll|ccccc}
\tablecaption{Fisher Matrix 1-$\sigma$ errors on parameters. 
\label{tab:parameterstable} }
\tablehead{
\colhead{Parameter}  & \colhead{Fiducial Value} & \colhead{Error Quoted} & 
\colhead{200 LOS} & \colhead{1000 LOS} & \colhead{200 LOS} & \colhead{1000 LOS} & 
\colhead{1000 LOS} \\  
\colhead{} & \colhead{} & \colhead{} & \colhead{+ 0 SIM} & \colhead{+ 0 SIM} & 
\colhead{+ 200 SIM} & \colhead{+ 200 SIM} & \colhead{+ 500 SIM} } 
\startdata
$\gamma(r_\star = 2 \, r_{\rm king})$ & 1.3 (0.32) & $\Delta \gamma_\star$  & 2.0 (2.2) & 0.91 (1.0) & 0.28 (0.24) & 0.21 (0.23) & 0.16 (0.15)\\
$M(r_\star = 2 \, r_{\rm king})$ [$10^7 {\rm M}_\odot$]  & 1.3 (5.3) & $\Delta M_\star/M_\star$ & 0.27 (0.38) & 0.12 (0.17) & 0.11 (0.11) & 0.07 (0.08) & 0.06 (0.05)   \\
$\rho(r_\star = 2 \, r_{\rm king})$ [$10^7 {\rm M}_\odot$ kpc$^{-3}$] & 4.0 (2.5) & $\Delta \rho_\star/\rho_\star$ &  1.6 (0.73) & 0.73 (0.33) & 0.16 (0.16) & 0.13 (0.12) & 0.09 (0.09) \\
$V_{\rm max}$ [${\rm km} \, {\rm s}^{-1}$] & 22 (28) & $\Delta V_{\rm max}/V_{\rm max}$ & 1.6 (1.7) & 0.71 (0.76) & 0.68 (0.93) & 0.45 (0.62) & 0.38 (0.52)\\
$\beta$ & 0 (0) & $\Delta \beta$  & 2.1 (1.4) & 0.97 (0.62) & 0.16 (0.16) &  0.15 (0.15)& 0.10 (0.10)\\
$a$ & 1 (0) & $\Delta a$ & 4.4 (4.6) & 2.0 (2.1) & 0.78 (0.87) & 0.55 (0.66) & 0.44 (0.51)\\
$\rho_0$ [$10^7 {\rm M}_\odot$ kpc$^{-3}$] & 1.0 (3.2)  & $\Delta \rho_0/\rho_0$ & 20 (9.2) & 9.1 (4.2) & 5.9 (2.0) & 4.0 (1.5) & 3.3 (1.1)\\
$r_0$ [kpc] & 2.0 (1.5) & $\Delta r_0/r_0$ & 8.3 (5.1) & 3.8 (2.3) & 3.3 (1.8) & 2.2 (1.2) & 1.8 (1.0)\\
\enddata
\tablecomments{Errors refer to the $1-\sigma$ range obtained from a full marginalization 
over the 6 input parameters ($a,b,c,r_0,\rho_0,\beta$).  
Quantities without (with) parenthesis refer to the cusp (core) model described in the text.
In the last 5 columns, LOS refers to the number of line-of-sight stars used to derive the quoted errors, 
and SIM refers to the number proper motions. 
}
\end{deluxetable*}

We pick large radial bins to compute the velocity dispersions and
  check that this uncorrelates the different bins.
Then the  elements of ${\bf F}$ are given by
\begin{eqnarray}
F_{\imath \jmath} = \sum_{M,\ell}
{1 \over \epsilon_{M\ell}^2} 
{\partial \sigma_{M\ell}^2 \over \partial p_\imath}
{\partial \sigma_{M\ell}^2 \over \partial p_\jmath } \, .
\label{eq:fishergaussian}
\end{eqnarray}
The sum is over $\ell$ radial bins and $M$ refers to
the three velocity ``methods'' --  one line-of-sight 
and two components in the plane of the sky.
The errors on the velocity dispersion are represented by
$\epsilon_{M\ell}$.
We choose bins of equal width in distance,  
so that there are approximately an equal number of stars in each
radial bin. 
As long as we distribute equal number of stars in each bin, the
results we present below are insensitive to the binning scheme, except  
in the limit of very few bins, or in the limit of small numbers of
stars per bin. 

To model the errors on the velocity dispersions, we define
$\epsilon_{M\ell}^2 
=\langle [ \sigma_{M\ell}^2 - \langle \sigma_{M\ell}^2 \rangle ]^2
\rangle$.
We assume that the errors on the velocity of each star are Gaussian and
that the theory error (from the distribution function) and
experimental error are summed in quadrature, 
\begin{equation}
\epsilon_{M\ell}^2 = \frac{2(n-2)}{n^2} 
\left [ {\sigma^t_{M\ell}}^2 +
  {\sigma^m_{M\ell}}^2 \right ]^2 \,,
\label{eq:totalerror} 
\end{equation}
where $n$ is the number of stars in each bin (taken here to be
constant from bin to bin). 
Here $\sigma^t_\ell$ is the dispersion in each bin determined from  
Eqs.~(\ref{eq:LOSdispersion})-(\ref{eq:phidispersion}), and
$\sigma^m_{M\ell}$  represents the measurement error in the velocities
of stars in that bin. 
For LOS velocities, the spectroscopic resolution implies errors 
$\sigma^m_{los} \sim 1 \, {\rm km} \, {\rm s}^{-1}$, 
which are negligible compared to the underlying dispersion, $\sigma^t \sim \, 10 \, {\rm km} \, {\rm s}^{-1}$.
 SIM can achieve $\sim 10 \, \mu$as yr$^{-1}$ proper motions for faint
stars
\footnote{
http://planetquest.jpl.nasa.gov/SIM/sim\_AstroIndex.cfm}.
At 100 kpc, this translates to an accuracy of $\sim 5$ km s$^{-1}$.
We will take this as a typical measurement error for velocities in the plane of
the sky.

The derivatives in  Eq.~(\ref{eq:fishergaussian}), and thus the errors
attainable on any of  the parameters,  depend on  the location  in the
true  set  of parameter space.   To examine  how the  errors vary as a
function of the true set of parameters, we choose two fiducial models.
Both  models produce a  ``typical''  dSph velocity dispersion profile,
which is roughly flat at $\sim 10 \, {\rm  km} \, {\rm s}^{-1}$ out to
$r_t$.  This requirement does not completely fix the profile.  For our
{\em  cusp} model (NFW) we add  the further restriction that $r_0$ and
$\rho_0$ fall within the expected CDM range
\citep{strigari_etal07}.   For our {\em  core}  model (Burkert) we set
$r_0$ and $\rho_0$ to give   a central phase space  density
$Q  \simeq 10^{-5}  \, {\rm M}_\odot \,
{\rm pc}^{-3} ({\rm km} \, {\rm s}^{-1})^{-3}$, which can be produced
in some WDM models \citep{Strigari:2006jf}.  
The values of $r_0$ and $\rho_0$ for both models are 
listed in  Table~\ref{tab:parameterstable} as are the implied log-slopes
$\gamma (r_\star) = 1.3$ (cusp) and $0.32$ (core) at the characteristic radius
$r_\star$, which is set below at the value $2 \, r_{\rm king} = 0.36$ kpc.

\begin{figure}
\plotone{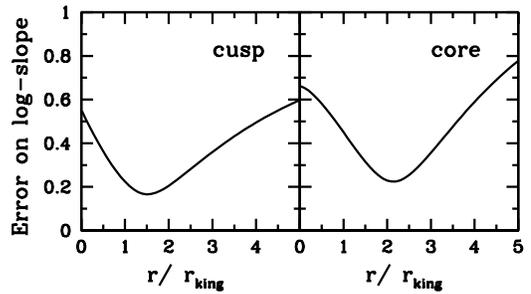}
\caption{\label{fig:gammastar} 
The 1-$\sigma$ error on the log-slope of the density 
profile as a function of the radius for 1000 LOS velocities
combined with 200 proper motions. The {\em left} and
{\em right} panels show our results for the cusp and
core cases, respectively.  The value of $r_\star$ is set to be where the error
is minimized.
 }
\end{figure}

\section{Constraining the Log-Slope} 
The log-slope of the density profile varies with radius, 
and so does the minimal error attainable on it. 
We are interested in searching for the radius, $r_\star$, where
the log-slope is
best constrained.
Fig.~\ref{fig:gammastar} shows the $1-\sigma$  error on $\gamma (r)$
computed   as  a  function  of  radius  using 1000
line-of-sight velocities and 200   proper motions for the cusp  (left)
and core (right) models.  In  both cases, the  error reaches a minimum
of $\Delta  \gamma \simeq 0.2$ at  $\sim 2 \, r_{\rm king}$ and
we explicitly adopt $r_\star = 2 \, r_{\rm king}$ for the rest of the
discussion.
We  find that  the value  of $r_\star$ and   the minimum error  depend
somewhat on the fiducial model.
As shown in Fig.~\ref{fig:gammastar}, profiles with cusps
are slightly  better  constrained   than  profiles  with cores,    and
$r_\star$ occurs at slightly smaller radii for cusped models.
We also find that as $r_t/r_{\rm king}$ increases, the
minimum   error      decreases       and    $r_\star$       increases.
We have chosen a conservatively small ratio here, $r_t/r_{\rm king} \simeq 5$.

Table~\ref{tab:parameterstable} summarizes our results.
We list 1-$\sigma$ errors
attainable on several quantities determined by marginalizing over the 6 fit
parameters discussed above.  We present errors for five  different combinations for the number
of LOS stars and proper motion stars.
  In all cases,  the errors on the ``physical''
parameters, $M (r_\star)$, $\rho(r_\star)$, and $\gamma (r_\star)$, decrease significantly when 
proper motions are included.
Quantities listed without (with) parentheses correspond to the cusp
(core) halo case. 
Note that the errors on the halo shape parameters $b$ and $c$ are
essentially the same as the priors we set on them of $\pm 2$ and hence
we do not list them in the table.

A quantity that is well  constrained without the addition of proper
motion information is the mass $M(r_\star)$, which can be determined
to $27 \%$ ($38 \%$) 
for cusped (cored) models with current data.  The  mass can be determined to
a remarkable $7-8 \%$ accuracy when 200 SIM stars are added.
The maximum circular
velocity is relatively unconstrained  by   LOS data alone,   but   becomes
determined to within $45\%(62\%)$ for cusped (cored) models with $200$ proper motions.
More interesting for the nature of dark
matter is the log-slope of the density profile at $r_\star$.  Even with
1000 LOS velocities, the log-slope in the cusp case is virtually unconstrained,
but with the addition of $200$ SIM stars the
log-slope is well-determined, e.g. $\gamma (r_\star) = 1.3 \pm 0.2$ for
the case of an underlying cusp.

The power of velocity dispersions to constrain the density and mass at
$r_\star \sim r_{\rm king}$ is relatively simple to understand
by examining the observable velocity dispersion components 
(Eqs.~\ref{eq:LOSdispersion},~\ref{eq:Rdispersion} and
\ref{eq:phidispersion}).
These observable quantities depend on the
  three-dimensional   stellar velocity    dispersion, which
scales as $\sigma_r^2(r) = \nu_\star^{-1}  r^{-2 \beta} \int_r^{r_t} G
\nu_\star(r) M(r) r^{2 \beta - 2} {\rm d}r \propto r^{2 -
  \gamma_\star}$ for a power-law stellar distribution $\nu_\star(r)$
and constant $\beta$.  The majority of stars reside at projected radii
$r_{\rm king} \lesssim R \lesssim r_{\rm t}$, where the stellar
distribution is falling rapidly $\nu_\star \sim r^{-3.5}$.  In this
case, for  $\beta = 0$, the LOS component scales as $\sigma_{los}^2(R)
\propto \int_R^{r_t} r^{-0.5 - \gamma_\star} (r^2 - R^2)^{-1/2} {\rm
  d}r$ and is dominated by the mass and density profile at the
smallest relevant radii, $r \sim r_{\rm king}$. For $R \lesssim r_{\rm
  king}$, $\nu_\star \propto r^{-1}$ and $\sigma_{los}^2$ is similarly
dominated by $r \sim r_{\rm   king}$ contributions.  Therefore we
expect the strongest constraints on the dark matter profile at
$r_\star \sim r_{\rm king}$.  For $\beta \ne 0$ similar arguments hold
but there is a manifest degeneracy between $\beta$ and the log-slope
$\gamma_\star$.  As discussed above and illustrated in Table 1, this
degeneracy can be broken by including both LOS information
(Eq. \ref{eq:LOSdispersion}) with tangential   information
(Eqs. \ref{eq:Rdispersion},~\ref{eq:phidispersion}). 

\begin{figure}
\plotone{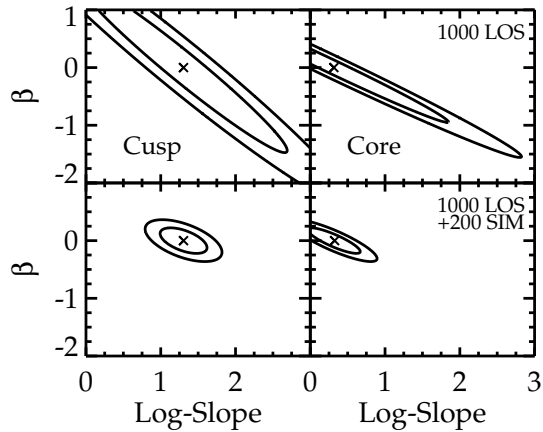}
\caption{\label{fig:contours} 
The $68\%$ and $95\%$ confidence regions for the dark halo density profile
slope measured at $r_\star = 2 r_{\rm king}$ and velocity anisotropy $\beta$.
The left (right) panels correspond to the cusp (core) halo model and
the small x's indicate the fiducial values.
Upper panels show the errors with 1000 LOS velocities
and the bottom panels show the errors for an additional 200 proper motions.
 }
\end{figure}

With the constraints on $\gamma_\star$, are we able to distinguish
cored from cusped models? To answer this question, in
Figure~\ref{fig:contours} we present $68\%$ and $95\%$ error contours
in the two-dimensional plane of $\gamma_\star$-$\beta$.  This figure
clearly shows the utility of combining both LOS and proper motions.
Due to the degeneracy with $\beta$, a sample of $\sim 1000$
line-of-sight stars is unable to distinguish between our two 
fiducial models, though the models can be clearly  distinguished with
the addition of proper motions. In the $\beta$-$\gamma_\star$ plane,
the principal constraining power comes from combining the LOS and $R$
velocity dispersions because the $\beta$ dependence in these two quantities
comes with opposite signs (Eqs. \ref{eq:LOSdispersion},~\ref{eq:Rdispersion}).

Finally, we address the fact that
$\beta$ could in principle vary significantly with radius. We have repeated our 
analysis assuming that
$\beta(r)=\beta_0+\beta_1r^2/(r^2+r_\beta^2)$,  and marginalize
over $\beta_0$, $\beta_1$ and $r_\beta$. We find very small changes to
the $\gamma(r_\star)$ error.  
For reasons similar to those given above, the three different velocity dispersion profiles
constrain $\beta(r=r_\star)$ to high accuracy, breaking the
$\beta$-$\gamma$ degeneracy.

\section{Summary} 
We have shown that the measurement of proper motions for
$\sim 200$ stars in a typical dSph can be combined with current 
line-of-sight velocity measurements to
constrain the dark halo log-slope to $\pm 0.2$ and normalization to $\pm 15 \%$
at about twice the King radius.
This is a
factor of $\sim 5$ better than currently possible with LOS velocity dispersion
data alone.
The results from such observations will provide a very 
sensitive test of the CDM paradigm and an incisive tool for investigating the
microscopic nature of dark matter.
We estimate that with $\sim 100$ days of observing time over 
the 5 year lifetime of SIM, it will be possible to obtain
the proper motions of $\sim 200$ stars in multiple dSphs \citep{simlong}. 

\section{Acknowledgments} 
We thank S. Kazantzidis, S. Kulkarni, S. Majewski, and R. Munoz for useful discussions.
LES is supported in part by a Gary McCue postdoctoral fellowship
through the Center for  Cosmology at the University of California,
Irvine. JSB, LES, and MK are supported in part by NSF grant
AST-0607746.

\clearpage

\clearpage

\clearpage

\end{document}